\documentclass[acmsmall]{acmart}

\AtBeginDocument{%
  }

\usepackage{array}

\renewcommand\footnotetextcopyrightpermission[1]{}
\begin{document}

\title{What is “Spatial” about Spatial Computing?}

\author{Yibo Wang}
\email{yibo828.wang@connect.polyu.hk}
\affiliation{%
  \institution{The Hong Kong Polytechnic University}
  \country{Hong Kong SAR, China}
}

\author{Yuhan Luo}
\email{yuhanluo@cityu.edu.hk}
\affiliation{%
 \institution{City University of Hong Kong}
 \country{Hong Kong SAR, China}}

\author{Janghee Cho}
\email{jcho@nus.edu.sg}
\affiliation{%
  \institution{National University of Singapore}
  \country{Singapore}
  }

\author{Junnan Yu}
\authornote{Corresponding Author}
\email{junnan.yu@polyu.edu.hk}
\affiliation{%
  \institution{The Hong Kong Polytechnic University}
  \country{Hong Kong SAR, China}
}

\begin{abstract}
 Recent advancements in geographic information systems and mixed reality technologies have positioned spatial computing as a transformative paradigm in computational science. However, the field remains conceptually fragmented, with diverse interpretations across disciplines like Human-Computer Interaction, Geographic Information Science, and Computer Science, which hinders a comprehensive understanding of spatial computing and poses challenges for its coherent advancement and interdisciplinary integration. In this paper, we trace the origins and historical evolution of spatial computing and examine how “spatial” is understood, identifying two schools of thought: “spatial” as the contextual understanding of space, where spatial data guides interaction in the physical world; and “spatial” as a mixed space for interaction, emphasizing the seamless integration of physical and digital environments to enable embodied engagement. By synthesizing these perspectives, we propose spatial computing as a computational paradigm that redefines the interplay between environment, computation, and human experience, offering a holistic lens to enhance its conceptual clarity and inspire future technological innovations that support meaningful interactions with and shaping of environments.
\end{abstract}



\maketitle
\section{Introduction}\label{sec:Introduction}
Nearly 170 years ago, during London’s cholera outbreak, Dr. John Snow pioneered the integration of spatial data by mapping cases relative to roads, property boundaries, and water pumps—ultimately tracing the disease to a single contaminated pump \cite{snow1856cholera}. This foundational work marked the first documented use of spatial analysis to link “what” and “where,” laying the groundwork for what is now known as spatial computing. Since then, the ability to harness spatial data—information embedded with locational attributes across geographic and non-geographic contexts~\cite{greenwoldSpatialComputing2003}—has expanded far beyond public health. Evolving from geographic information science (GIScience), spatial computing now integrates artificial intelligence, sensor technologies, and human-computer interaction \cite{shekharSpatialComputing2020} to power applications ranging from urban planning and GPS navigation to autonomous vehicles and humanoid robots \cite{xieTransformingSmartCities2018, jungRevolutionizingElectricVehicle2022, fernandez-rojas_contextual_2019, rozlivek_harmonioushuman-like_2025}. In the past decade, immersive technologies like mixed reality (MR) have further broadened its scope, blending digital and physical spaces to create lifelike hybrid environments. With public interest surging—rising 5,500\% after the Apple Vision Pro launch in 2024~\cite{LinkedIn2024}—and a market projected to grow from USD 97.9 billion in 2023 to USD 280 billion by 2028 \cite{FutureMarketInsights2022}, spatial computing is rapidly becoming a transformative force in both technology and everyday life.

Despite the growing popularity and interest in spatial computing, there remains a lack of a comprehensive and systematic understanding of the concept. Researchers from different disciplines such as Human-Computer Interaction (HCI), GIScience, Architecture, and Computer Science offer varying perspectives on its framing and characteristics. For instance, some define spatial computing as a “technology” or “system” that integrates virtual objects and environments within a physical space context \cite{keshavarziContextualSpatialComputing2022}. In contrast, others argue that spatial computing should be viewed as an interaction paradigm, describing it as “a concept that describes the way in which humans communicate with our digital landscape” \cite[p.~212]{gargSpatialComputingNext2023}. Such conceptual ambiguity poses significant challenges to the effective advancement of the field, particularly in technical and HCI research on spatial computing, as it hinders the necessary theoretical foundation needed to unify the field and guide innovation while also complicating system development by preventing the establishment of consistent evaluation criteria for interfaces, interaction paradigms, and computational architectures \cite{olson_concepts_2014}. Therefore, critically examining and clarifying the conceptual and practical aspects of spatial computing is essential to building a solid foundation for its continued evolution.

In this paper, we aim to bring conceptual clarity to spatial computing by tracing its historical origins and development, analyzing its key characteristics, and identifying two distinct schools of thought regarding the meaning of “spatial”: (1) as the contextual understanding of space, where spatial data guides interactions in the physical world, and (2) as a mixed space for interaction, seamlessly integrating physical and digital spaces for embodied engagement. We emphasize space as an essential computational component in today’s technological landscape and outline key future research directions. Rather than proposing rigid categories or a singular definition, we acknowledge that the meaning of spatial computing varies with context, adapting to diverse disciplinary perspectives and needs. We hope this paper can offer conceptual tools and practical insights to help researchers and practitioners navigate its complexities, recognizing spatial computing as a transformative paradigm for engaging with space in the digital era. The subsequent sections provide a comprehensive overview of its historical development, present the two schools of thought, discuss related applications and challenges, and conclude by reflecting on the field’s conceptual foundations and future research directions.

\section{Origin and Evolution of Spatial Computing}\label{sec:History}
\begin{figure}[ht]
    \centering
    \includegraphics[width=\textwidth]{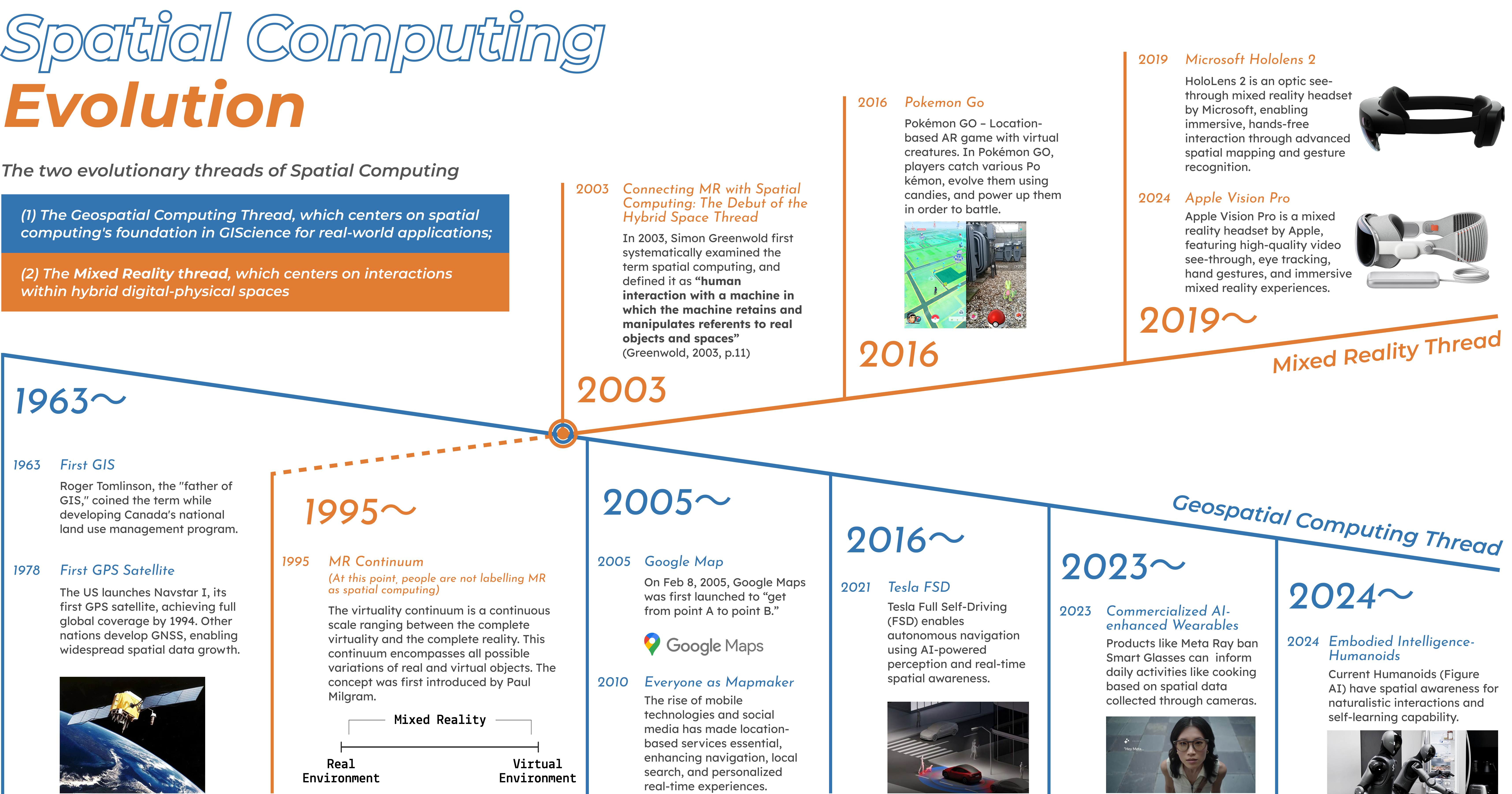}
    \caption{The Origin of Spatial Computing and Its Two Evolutionary Threads: The Geospatial Computing and Mixed-Reality threads}
    \label{fig1}
\end{figure}

The term “spatial computing” was first introduced by Simon Greenwold in 2003 in the context of virtual environments and computational modeling~\cite{greenwoldSpatialComputing2003}. This early framing has significantly shaped how researchers perceive the field today, often linking it with immersive technologies such as augmented reality (AR) and virtual reality (VR)~\cite{keshavarziContextualSpatialComputing2022, ballestasFrameworkCentralizingEthics2021}. However, the technological foundations of spatial computing extend further back to 1963 with the development of the first geographic information system (GIS)~\cite{maguire1991overview}, which marked the beginning of integrating computational processes with spatial data to support meaningful interactions and services. For example, Shekhar et al. describe spatial computing as encompassing the ideas, tools, technologies, and systems that reshape our understanding of location—how we perceive, communicate, and navigate spatial relationships~\cite{shekhar_spatial_2015}. By examining the field’s historical development and technical evolution, we identify two primary threads of spatial computing (Figure~\ref{fig1}): (1) the Geospatial Computing Thread, grounded in GIScience and focused on physical-world applications, and (2) the Mixed Reality Thread, which emphasizes interaction within hybrid digital-physical environments.

\subsection{Geospatial Computing Thread}
The geospatial computing thread forms the historical foundation of spatial computing, originating in GIScience and focusing on the use of geographic information to enhance human-machine interactions in the physical world. Emerging in the 1960s with advances in geographic information technologies, early spatial computing centered on processing static geospatial data through tools like remote sensing and GIS to support public services such as urban planning, emergency response, and asset tracking \cite{adamSpatialComputingSocial2012}. Over time, it evolved to include spatiotemporal elements—like real-time GPS tracking—enabling dynamic analysis across space and time, and expanding applications to public health, safety, and agriculture \cite{shekharSpatialComputing2020}. Initially, spatial computing was centralized, complex, and predominantly managed by government and industry experts, which, along with technical limitations (e.g., low-resolution data, limited computing power~\cite{shekharGPSVirtualGlobes2015}), restricted public access. However, starting in the mid-1980s, improvements in GIScience and technologies like in-car GPS began to democratize its use. By around 2010, smartphones enabled billions to generate and share geospatial data, making spatial tools accessible beyond experts. Today, mobile technologies and social media have transformed spatial computing into a ubiquitous, user-centered paradigm, where individuals actively engage with location-based services in real-time \cite{adamSpatialComputingSocial2012}.

Beyond navigation-related services, contemporary geospatial-based spatial computing is evolving to support context-aware interactions by integrating environmental awareness. Embedded sensors, processors, and actuators have now been integrated into commercial products to capture and process room-scale spatial data. Such an emphasis on environmental awareness not only enhances daily activities but also expands the impact of spatial computing across various industries, such as healthcare and manufacturing. Furthermore, technological advancements in artificial intelligence (AI), light detection and ranging (LiDAR), and robotics are enabling machines to better perceive, map, and navigate their surroundings in real time, enhancing their autonomy for more precise object recognition, obstacle avoidance, and adaptive decision-making in dynamic environments. By processing real-time spatial data, these technologies support informed decision-making, precise movement, and adaptive, human-like intelligence. Together, these advancements have transformed geospatial-based spatial computing into an intelligent, interactive, and adaptive ecosystem—seamlessly integrating real-time data, environmental awareness, and human-machine collaboration to enhance everyday experiences and drive innovation across industries.

\subsection{Mixed Reality Thread}

In contrast, the mixed-reality (MR) thread of spatial computing focuses on human-computer interactions that occur across physical and virtual spaces. In the early 1990s, Milgram et al. \cite{milgram_augmented_1995} introduced the reality-virtuality continuum, which spans from a fully real environment (the physical world) to a fully virtual environment (virtual reality, or VR). The intermediate spectrum—excluding these two extremes—is classified as MR, highlighting its function as a conduit between physical and digital experiences. In 2003, Simon Greenwold integrated MR with the concept of spatial computing, defining spatial computing as \textit{“human interaction with a machine in which the machine retains and manipulates referents to real objects and spaces”} \cite[p.~11]{greenwoldSpatialComputing2003}. He argued that the development of hybrid experiences, such as MR, depends on machines’ ability to understand and incorporate physical space into computational systems, signifying the convergence of MR and spatial computing. This convergence expanded spatial computing’s scope beyond interactions driven solely by geospatial data in the physical world to encompass interactions within hybrid spaces \cite{xuSpatialComputingDefining2024}, thereby establishing the foundation of the mixed reality thread in spatial computing.

Over the past decade, advancements in immersive technologies have dissolved the boundaries between physical and virtual spaces, resulting in a growing prevalence of hybrid experiences. The contemporary mixed reality thread of spatial computing focuses on immersive technologies that accurately map real-world environments and seamlessly integrate them into digital contexts. This integration facilitates sophisticated interactions in which digital content not only coexists with but also modifies physical reality. As a result, more intuitive and adaptive user interfaces and interactions have emerged, enabling users to engage with hybrid spaces naturally through gestures, speech, and movement, primarily via head-mounted displays. Overall, the mixed reality thread has expanded spatial computing into an immersive, adaptive paradigm that seamlessly merges physical and digital environments to enhance communication, design, and interaction across domains.

In summary, spatial computing has advanced through machines’ growing ability to interpret and interact with the physical world, leading to two main paradigms: geospatial applications in physical space and immersive experiences in mixed reality. This focus on spatial awareness underscores the central role of space in shaping human-environment interaction and integrating digital information. The next section explores the varied meanings of “spatial” within this context.

\section{The Meaning of “Spatial” in Spatial Computing: Two Schools of Thought}\label{sec:Design}
Due to its diverse evolution, broad applications, and varying interpretations, spatial computing still lacks a unified understanding. Scholars and practitioners approach it from different angles: some view it as a transformative interaction paradigm that reshapes how we engage with spatial information~\cite{shekharSpatialComputing2020, ballestasFrameworkCentralizingEthics2021}, while others frame it as a technological framework that integrates digital and physical realities~\cite{keshavarziContextualSpatialComputing2022, brodersenSpatialComputingSpatial2007}. Yet, across these perspectives, a shared emphasis is \textbf{the central role of spatial experience mediated by computation.}  To build a clearer and more nuanced understanding, it is essential to examine how “spatial” is conceptualized within these frameworks. We identify two primary schools of thought: \textit{\textbf{“spatial” as the contextual understanding of space}} and \textbf{\textit{“spatial” as a mixed space for interaction}}. The perspectives are not limited to the evolutionary threads previously discussed but are evident across the field. The following subsections examine each interpretation of “spatial” and its related applications and challenges.

\subsection{“Spatial” as the Contextual Understanding of Space}

The first school of thought, \textit{“Spatial” as the Contextual Understanding of Space,} emphasizes the acquisition, representation, and utilization of spatial data to enable digital systems to interact meaningfully with the physical world. Initially, this perspective was primarily concerned with locating and positioning objects, as seen in early geospatial computing applications such as in-car GPS navigation systems. However, with advancements in geospatial analytics, sensor-driven spatial awareness, and structured data flows, spatial computing has evolved beyond static spatial representations into real-time, semantic spatial models~\cite{kuhnDesigningLanguageSpatial2015}. These models allow for the instant capture, analysis, and computational representation of physical environments, enabling applications such as autonomous navigation (e.g., self-driving cars) and human-environment interactions (e.g., location-based mixed-reality games like Pokémon Go).

To achieve a contextual understanding of space, spatial computing relies on \textbf{computational models that structure and interpret spatial environments.} These models—such as point clouds, 3D meshes, and semantic maps—transform raw spatial data into machine-readable representations, making space accessible, measurable, and manipulable. Key contextual data elements, such as location, fields, objects, networks, events, accuracy, and granularity, form the foundation of these computational models, allowing machines to derive meaningful spatial relationships and contextual awareness~\cite{kuhnDesigningLanguageSpatial2015}. By leveraging advanced analytical techniques like object recognition, scene segmentation, and spatial reasoning, these models support real-time decision-making in dynamic environments. For instance, autonomous vehicles must continuously analyze road conditions, pedestrian movements, and traffic signals to navigate safely. In this sense, spatial computing operates as a \textbf{data-driven paradigm} where high-quality spatial data is essential for enhancing a system’s ability to interpret and respond to its surroundings, ultimately improving spatial reasoning, environmental modeling, and navigation \cite{shekharSpatialComputing2020, ballestasFrameworkCentralizingEthics2021}.

Building on this foundation, spatial context awareness extends computational modeling by integrating the \textbf{social dynamics of human activities} into spatial computing. As computing infrastructures and data sources have evolved, spatial computing has transitioned from static mapping to dynamic, real-time spatial intelligence, allowing digital systems to adapt to human behaviors, environmental changes, and situational contexts. This evolution becomes a cornerstone for a wide range of fields, such as robotics \cite{delmericoSpatialComputingIntuitive2022}, where precise spatial modeling and analysis are critical for problem-solving and informed decision-making. By integrating structured spatial data with contextual factors, this approach enables \textbf{more adaptive, responsive, and human-aware }spatial computing applications, bridging the gap between digital and physical environments.

\subsubsection{Representative Applications}
In terms of applications, this contextual-understanding-based spatial computing paradigm encompasses a diverse range of use cases, broadly categorized into three types based on key technological features. The first category includes \textbf{GIScience-based applications}, which leverage locating and positioning technologies such as GIS, GPS, remote sensing, and spatial data science. These applications facilitate urban infrastructure development, smart traffic management, and the optimization of public services, including disaster response and public transport coordination \cite{adamSpatialComputingSocial2012}. Additionally, GIScience underpins navigation and location-based services, powering real-time GPS-based navigation systems like Google Maps. The second category involves \textbf{spatial mapping-based applications}, which drive advancements across various fields. In architecture and built environments, MR-assisted design and digital twin models enable real-time urban planning and simulation \cite{weberSpatialComputingBuilding2024}. In healthcare, spatial mapping enhances medical imaging visualization, AR-assisted surgery, and rehabilitation through interactive MR applications. Additionally, autonomous robotics benefits from spatial mapping in applications such as AI-driven robotic automation and remote robotic-assisted procedures, particularly in industrial and medical settings. The third category, \textbf{AI-driven and sensor fusion-based spatial computing applications}, fuels innovations in embodied intelligence, enabling humanoid robots with spatial awareness for human-like interactions. AI also enhances wearable computing, exemplified by Meta Ray-Ban Smart Glasses, which integrate real-time AI-driven voice assistance. Furthermore, autonomous mobility solutions, such as Tesla’s Full Self-Driving (FSD) system, utilize AI-driven spatial computing for autonomous navigation and vehicle automation. Collectively, these advancements enable machines to interpret and respond to spatial environments in real time, improving decision-making and adaptive human-computer interaction across domains.

\subsubsection{Major Challenges}

Despite its transformative applications, the spatial data-driven approach to spatial computing faces some fundamental challenges. A primary issue is \textbf{the lack of semantic clarity in spatial data. }Traditionally, GIS systems have prioritized syntactic structure over semantic meaning, resulting in limited frameworks for organizing and interpreting contextual spatial information \cite{kuhnDesigningLanguageSpatial2015}. This hinders spatial computing systems’ ability to process and extract meaningful insights, reducing their effectiveness across applications. Therefore, achieving precise real-time contextual awareness remains difficult. While spatial computing excels at capturing geometric structures, it struggles with semantic recognition, particularly in identifying the function of spaces and their dynamic transformations over time. Another critical challenge involves \textbf{ data privacy and ethical concerns}, as spatial computing systems rely on continuous environmental sensing and monitoring, often collecting personal and bystander information \cite{keshavarziSynthesizingNovelSpaces2022}. These datasets may include biometric details, movement patterns, and interaction histories, raising concerns about surveillance, unauthorized data usage, and potential privacy breaches. Additionally, many spatial data systems \textbf{require high computational resources}, making them costly for small enterprises and educational institutions, while the \textbf{lack of cross-disciplinary integration} in spatial computing—characterized by fragmented solutions that hinder seamless interaction between GIS, AI, and sensor-based systems—further contributes to inefficiencies in its applications. Advancing semantic data interpretation and ethical governance is key to making spatial computing more robust, responsible, and broadly impactful across disciplines.

\subsection{“Spatial” as a Mixed Space for Interaction}
The second school of thought centers on the concept of \textit{“Spatial” as a Mixed Space for Interaction,} framing space as an interactive, computationally rich medium where users engage with digital systems through spatially embedded interactions rather than conventional interfaces. This perspective envisions the surrounding environment as an active interface, seamlessly integrating digital and physical elements, particularly through technologies such as augmented and mixed reality \cite{ballestasFrameworkCentralizingEthics2021, keshavarziContextualSpatialComputing2022}. Scholars and practitioners in HCI research have long advocated for this shift, arguing that computation should extend beyond screens into the physical world. As Simon Greenwold \cite[p.~8]{greenwoldSpatialComputing2003} asserted, “Computation’s denial of physicality has gone about as far as it can; it is time for a reclamation of space as a computational medium.” Such a shift suggests that human-computer interaction should become more intuitive, aligning with natural human perception and movement to foster immersive, responsive, and human-centered experiences.

At the core of this hybrid-space-based spatial computing is the concept of \textbf{data corporeality,} which challenges the traditional view of digital information as abstract by emphasizing its tangible presence and real-world implications \cite{ishii_tangible_1997}. Historically, as computers became more powerful, their physical footprint diminished, creating a tension between human spatial perception and the digital realm \cite{greenwoldSpatialComputing2003}. Spatial computing reconciles this tension by embedding computation into the physical environment, shifting digital interactions from screen-based interfaces to spatially integrated experiences. Unlike conventional computing paradigms that separate physical and digital realms, \textbf{spatial computing treats space itself as a computational medium,} enabling seamless, embodied interactions that merge virtual and real elements. This approach, intertwined with ubiquitous computing, tangible interfaces, and invisible computing \cite{weiserComputer21stCentury1999, norman1998invisible}, transforms machines from passive tools to active participants in shaping spatial interactions, fostering more intuitive and dynamic engagements between humans, technology, and the environment. 

As digital interaction shifts to spatially integrated experiences, spatial computing redefines user engagement with computational systems, giving rise to the concept of the \textbf{interfaceless experience}, where digital artifacts function as adaptive systems that respond dynamically to user actions and environmental factors in real time. Enabled by AI-driven spatial contextual awareness, this shift allows for personalized, responsive interactions, eliminating traditional user interface (UI) constraints and making digital engagement more natural and human-centric. Additionally, spatial computing introduces new sensory engagement modes, integrating vision, sound, touch, and movement to enhance immersion. Technologies like hand tracking, spatialized audio, and AI-driven eye tracking enable users to interact with digital elements as seamlessly as with physical objects, expanding the expressive range of human-computer interaction \cite{pangilinanCreatingAugmentedVirtual2019}. This transformation positions computation not merely as a tool for executing commands but as an active participant in shaping human experience within spatial environments.

\subsubsection{Representative Applications}
This mixed space paradigm of spatial computing integrates computer vision, sensor fusion, and MR interfaces to seamlessly blend digital and physical environments, enabling immersive and adaptive interactions. Currently, relevant applications primarily fall into three main categories: Mobile Augmented Reality (AR), Optic See-Through MR, and Video See-Through MR. \textbf{Mobile AR applications}, such as Pokémon GO, Instagram AR Filters, and Google Live View, deliver AR experiences through mobile devices using cameras and sensors. \textbf{Optic See-Through MR}, used in devices like Microsoft HoloLens 2 and Magic Leap 2 employs transparent lenses for digital overlay. \textbf{Video See-Through MR}, as seen in Apple Vision Pro and Meta Quest 3, captures real-world views and overlays digital content, facilitating interaction with virtual elements in real-world settings. These applications are widely used across diverse fields, such as design (e.g., MR architecture), manufacturing (e.g., AR glasses for productivity; teleoperating robotics), education (e.g., MR collaboration), entertainment (e.g., Pokémon GO), and healthcare (e.g., rehabilitation simulations).

\subsubsection{Major Challenges}

Leveraging space as an interactive medium faces three main challenges: interaction design, social disruption, and environmental constraints. First, \textbf{achieving intuitive interfaceless interactions is critical}, as the absence of standardized design guidelines for gesture recognition, spatial mapping, and real-time contextual adaptation often leads to inconsistent and frustrating user experiences in immersive environments. The shift from 2D interfaces to 3D spatial interactions can cause decision paralysis due to excessive freedom in environments lacking clear affordances, while over-automation in AI-driven systems risks overriding user intent by making autonomous decisions without sufficient transparency or user control. Second, \textbf{mixed spaces can disrupt social norms}, complicating communication as augmented content may interfere with conversations, obscure nonverbal cues, and introduce distractions \cite{ballestasFrameworkCentralizingEthics2021}. AR glasses and MR headsets blur the line between digital and physical spaces, creating asymmetrical information dynamics where users access augmented information that bystanders cannot perceive \cite{sethumadhavanFrameworkEvaluatingSocial2021}. Finally, spatial computing must \textbf{contend with real-world environmental constraints}, such as lighting variations, occlusions, and space limitations, which impede object recognition and MR interactions \cite{keshavarziContextualSpatialComputing2022}. Addressing these challenges requires balancing automation with user agency and developing standardized design frameworks to ensure intuitive and context-aware spatial interactions.

\begin{table}[!ht]
\caption{Two Perspectives of Spatial Computing and Their Features, Applications, and Challenges}
\label{tab:spatial_paradigms}
\centering
\scriptsize
\renewcommand{\arraystretch}{1.2}
\begin{tabular}{>{\raggedright\arraybackslash}m{2cm} >{\raggedright\arraybackslash}m{3cm} >{\raggedright\arraybackslash}m{2.5cm} >{\raggedright\arraybackslash}m{2.5cm} >{\raggedright\arraybackslash}m{2.2cm}} 
\toprule
\textbf{Perspective} & \textbf{Brief Explanation} & \textbf{Technological Features} & \textbf{Application Areas \& Cases} & \textbf{Challenges} \\
\midrule
Contextual Understanding of Space
& Capturing, structuring, and analyzing spatial data; emphasizes computational representation of space for machine interpretation.
& Utilizes GIS, GPS, remote sensing, spatial databases, and AI-powered spatial analytics.
& Geoscience, urban planning, disaster management, autonomous navigation, embodied intelligence, AI-driven wearable computing, etc.
& Insufficient semantic understanding of space, data privacy concerns, accessibility barriers, and the need for interdisciplinary frameworks. \\
\midrule
A Mixed Space for Interaction
& Conceptualizes space as an interactive environment where digital and physical elements co-exist; prioritizes interfaceless, immersive interactions.
& Leverages AI, sensor fusion, XR, tangible interfaces, and adaptive multimodal interaction techniques.
& Architecture, gaming, robotics, healthcare, and human-machine collaboration, etc.
& Insufficient real-time context awareness, over-automation, social acceptability, and interaction design. \\
\bottomrule
\end{tabular}
\end{table}

\section{Closing Reflections}\label{sec:Discussions}
In this article, we explore the origins and historical development of spatial computing, identifying two evolutionary threads based on technological origins (i.e., interaction driven by geospatial computing and interaction within mixed reality) and two schools of thought regarding the interpretation of “spatial” -- as the contextual understanding of space and as a mixed space for interaction. We examine each perspective’s core concepts, technological dimensions, application areas, and the associated challenges (see Table~\ref{tab:spatial_paradigms}). It is important to note that these two perspectives are not mutually exclusive, as many recent applications integrate elements of both. For instance, AI-enhanced wearables like the Apple Vision Pro leverage precise spatial mapping and sensor fusion to understand real-world environments (i.e., contextual understanding of space) while enabling context-aware, multi-sensory interactions through eye tracking, hand gestures, and MR overlays (i.e., mixed space for interaction). In this section, we further discuss how our synthesis advances the conceptual and theoretical foundations of spatial computing for its ongoing evolution.

\subsection{Understanding Spatial Computing as an Integrative Computational Paradigm}
Our synthesis reveals significant variation in how spatial computing is conceptualized across disciplines. While existing definitions address important aspects of spatial computing (e.g., as a broad category of GIScience technologies~\cite{shekharGPSVirtualGlobes2015}, a mode of human-computer interaction~\cite{greenwoldSpatialComputing2003}, or a mixed reality framework), they are often fragmented and incomplete, frequently neglecting its broader theoretical, technological, and application-oriented implications. Unlike traditional computing paradigms that treat space as a passive container for computation and interaction~\cite{weiserComputer21stCentury1999}, spatial computing actively structures, interprets, and interacts with spatial environments. This capability arises from two interrelated dimensions. First, it emerges from the convergence of multiple disciplines, integrating insights from GIScience, HCI, AI, robotics, and architecture into a unified framework. For example, spatial computing systems leverage geospatial analytics from GIScience to interpret environments \cite{shekharGPSVirtualGlobes2015}, incorporate principles of embodied interaction and natural user interfaces from HCI for intuitive interactions \cite{ishii_tangible_1997, dourish_where_2001}, and apply machine learning techniques from AI to dynamically adapt to surroundings \cite{lecunDeepLearning2015}. Second, it synthesizes two fundamental conceptualizations of “spatial”: spatial as contextual understanding, where computation derives meaning from environmental cues, and spatial as a mixed space, where digital and physical environments seamlessly integrate to enable embodied interaction. Together, these dimensions define the essence of spatial computing as a multi-dimensional, integrative paradigm that extends beyond technical implementations to shape human experiences in computationally enriched spaces.

By integrating both disciplinary perspectives and conceptual understandings of space, spatial computing transcends its characterization as a mere technological advancement, instead emerging as a framework for designing intelligent, adaptive environments in which computation and space co-evolve to support human activity. As such, we propose that \textit{\textbf{spatial computing should be understood as an integrative computational paradigm that redefines the relationship between space, computation, and human experience, transcending isolated technical advancements to enable the seamless convergence of digital and physical environments into context-aware, adaptive, and interactive systems that enhance human experiences.}} This integrated approach highlights spatial computing’s multidimensional complexity, spanning theoretical, technological, and interactional domains, bridging disciplinary divides, and creating opportunities for the development of intelligent, spatially aware systems that dynamically adapt to human needs and environmental contexts. For instance, navigation systems in AR glasses integrate GIS mapping, AI-driven contextual awareness \cite{guptaCognitiveMappingPlanning2017}, and HCI principles \cite{norman1998invisible, pangilinanCreatingAugmentedVirtual2019} to provide real-time, location-based guidance that overlays digital information onto the physical world. Overall, by adopting an integrative perspective that embraces both its interdisciplinary foundations and its dual spatial conceptualizations, we can move toward a more cohesive, theoretically grounded, and practically impactful understanding of spatial computing.

\subsection{Space as an Increasingly Important Computational Component}

In our exploration of the term “spatial” within spatial computing, we have identified two key perspectives that shape its conceptual and technological framework: Contextual Understanding of Space and Mixed Space for Interaction. Our identification of these two perspectives on space offers, for the first time, a structured and systematic approach to integrating diverse perspectives on spatial computing, providing a unified foundation for the field’s future effective research and development. Although the two perspectives focus on different aspects of space (i.e., contextual data versus interaction medium), both underscore the essential role of space in shaping technology-mediated human experiences. In the Contextual Understanding perspective, effective human-machine interaction relies on a system’s ability to interpret and respond to environmental contexts for adaptive decision-making. Meanwhile, the Mixed Space perspective views space as a computationally enriched medium, where digital augmentation enhances perception, embodiment, and agency in hybrid environments~\cite{blanchardUseMixedReality2022, sasiFutureInnovationHealthcare2021, akersMixedRealitySpatial2020a}. Crucially, \textbf{both perspectives challenge the traditional notion of space as a passive backdrop for computation, instead positioning it as an active, structured component that informs machine intelligence and shapes human interaction.} 

By structuring \textbf{space as an active computational variable}, systems can dynamically adapt to environmental changes, enhancing context-aware AI, real-time decision-making, and spatial reasoning in spatial computing applications like robotics and autonomous systems. For example, in smart environments, this perspective enables responsive infrastructures, such as buildings that adjust to occupancy patterns or urban systems that optimize traffic flow using real-time spatial analytics~\cite{shekhar_spatial_2015}. Similarly, in AR/VR, spatial properties can be computationally manipulated to enhance user engagement, navigation, and interaction to enable more immersive and intuitive interfaces~\cite{delmericoSpatialComputingIntuitive2022}. However, many spatial computing systems often prioritize technological implementation—sensor accuracy, algorithmic efficiency, or hardware performance—without fully integrating space as an interactive component that shapes human cognition and behavior, limiting their potential for truly intelligent, human-centered design. To advance spatial computing applications, researchers and practitioners must go beyond technical optimization and ensure that systems effectively incorporate spatial relationships, human cognition, and real-world dynamics. For example, how can these systems seamlessly integrate human perception, digital augmentation, and spatial context? How can they enhance user agency and embodiment while ensuring natural, intuitive interactions~\cite{tangUserExperienceTechnical2024}? By adopting a spatially conscious design approach, future innovations can create more adaptive, intuitive, and human-centered systems that align seamlessly with the way people think, move, and interact with their environments.

\subsection{Future Directions and Opportunities}
Recent advancements in AI, sensor technologies, and computational modeling have significantly enhanced the capabilities of spatial computing, with ongoing research poised to drive further progress. Across both perspectives, several key challenges emerge in the design and development of spatial computing technologies, such as semantic understanding, intuitive interaction design, and broader ethical considerations. Beyond the common challenges shared by many emerging technologies (e.g., ethics and accessibility), we especially highlight what we identify as the critical future directions for the continued evolution of spatial computing. 

\subsubsection{Semantic Spatial Understanding: From Positioning to Meaning}

Both the Contextual Understanding of Space and Mixed Space for Interaction perspectives emphasize the transformation of space from a static, mapped entity to a dynamically interpreted and computed environment (simulation environment). Traditional geospatial systems, such as GPS and GIS, organize physical space into structured data, primarily serving positioning and navigation needs. However, spatial computing has advanced beyond traditional mapping, evolving into real-time sensemaking (interpretation), where AI-driven perception infers meaningful information from contextual data sampled from physical space \cite{lecunDeepLearning2015}. The transition from non-semantic, simple spatial perception to semantic and complex spatial cognition marks a shift where space is not only quantified but also endowed with meaning based on its contextual and dynamic properties \cite{ZhaoPyramid, Zhang_2023_CVPR}. A primary challenge in this shift is enabling machines to understand spatial context rather than merely recognize and track positions. Current AI-driven applications, such as autonomous navigation, embodied AI, and smart environments, illustrate the growing importance of semantic-level spatial representation learning—machines must not only detect objects but also understand how space is used, how it changes over time, and how digital and physical spaces interact. Future research should therefore develop robust models of spatial semantic reasoning, integrating sensor fusion and real-time contextual reasoning. However, a key question remains: How can technologies develop a deeper, human-like understanding of spatial context to enhance interaction, adaptation, and decision-making? Addressing this is key to enabling the next generation of spatially intelligent systems in robotics, assistive tech, and interactive environments.

\subsubsection{Spatial Interaction Frameworks: Designing for Intuitive and Adaptive Engagement}
Spatial computing redefines the relationship between humans, computation, and space by turning the environment into an extension of digital cognition and interaction, shifting from screen-based interfaces to interfaceless, spatially embedded experiences. However, these interactions are not inherently intuitive and require new spatial literacies beyond traditional abilities to interpret spatial patterns and relationships~\cite{bednarz2011understanding}. As spatial computing blurs the boundaries between digital and physical, it demands fluid, context-aware interaction paradigms that move beyond standardized inputs like keyboards and mice, raising critical design questions such as how embedded digital experiences can remain accessible without visible interfaces and how systems can balance automation with human agency in adaptive environments. Addressing these questions requires new frameworks for intuitive spatial interaction design through empirical investigations with users, considering both cognitive ergonomics and sociocultural factors. Future research should develop scalable, inclusive, and intuitive interaction models, drawing on insights from proxemics to align interactions with social norms~\cite{ballendat2010proxemic}, and incorporating haptic feedback, multimodal AI, and adaptive systems to enhance user engagement. By bridging the gap between digital and physical spaces, spatial computing has the potential to create more seamless, immersive, and human-centered interactions, ultimately transforming how we engage with technology in our everyday lives.

\section*{Acknowledgment}
This work was funded by The Hong Kong Polytechnic University (\#: P0056314)


\bibliographystyle{ACM-Reference-Format}
\bibliography{bibliography}

\end{document}